\documentclass[letter]{aa}

\usepackage[utf8]{inputenc}
\usepackage[varg]{txfonts}
\usepackage{textcomp}      

\usepackage{natbib}
\bibpunct{(}{)}{;}{a}{}{,} 
\usepackage{graphicx}   
\usepackage{epstopdf}
\usepackage{hyperref}   
\usepackage{longtable}
\usepackage{amsfonts,amssymb,latexsym}
\usepackage[normalem]{ulem}
\usepackage[dvipsnames]{xcolor}

\hypersetup{
    bookmarks=true,    
    colorlinks=true,   
    linkcolor=blue,    
    citecolor=blue,    
    filecolor=blue,    
   urlcolor=blue       
}

\def\ga{\,\,\raise0.14em\hbox{$>$}\kern-0.76em\lower0.28em\hbox{$\sim$}\,\,}
\def\la{\,\,\raise0.14em\hbox{$<$}\kern-0.76em\lower0.28em\hbox{$\sim$}\,\,}

\def\Msun{~$M_{\odot}$}

\def\cm3{cm$^{-3}$}

\def\chem#1#2{$\mathrm{^{#2}\kern-0.8pt#1}$}
\def\reac#1#2#3#4#5#6{$\mathrm{\, ^{#2}\kern-0.8pt{#1}\, ({#3}\, ,{#4})\, {}^{#6}\kern-0.8pt{#5}\, }$}

\def\be{\begin{equation}} 
\def\ee{\end{equation}}
\def\beqy{\begin{eqnarray}}
\def\eeqy{\end{eqnarray}}
\def\bmlet{\begin{mathletters}}
\def\emlet{\end{mathletters}}

\begin{document}

\title{Proton ingestion in asymptotic giant branch stars as a possible explanation for J-type stars and AB2 grains}

\author{A. Choplin
\and 
L. Siess
\and
S. Goriely
}
\offprints{arthur.choplin@ulb.be}

\institute{
Institut d'Astronomie et d'Astrophysique, Universit\'e Libre de Bruxelles,  CP 226, B-1050 Brussels, Belgium
}

\date{Received --; accepted --}

\abstract
{
J-type stars are a subclass of carbon stars that are generally Li-rich, not enriched in s-elements, and have low $^{12}$C/$^{13}$C ratios. They were suggested to be the manufacturers of the pre-solar grains of type AB2 (having low $^{12}$C/$^{13}$C and supersolar $^{14}$N/$^{15}$N).   
}
{
In this Letter, we investigate the possibility that J-type stars are early asymptotic giant branch (AGB) stars that experienced a proton ingestion event (PIE).
}
{
We used the stellar evolution code \textsf{STAREVOL} to compute AGB stellar models with initial masses of 1, 2, and 3\Msun{} and metallicities   [Fe/H]~$= -0.5$ and 0.0. We included overshooting above the thermal pulse and used a network of 1160 nuclei coupled to the transport equations. The outputs of these models were compared to observations of J-type stars and AB2 grains.
}
{
In solar-metallicity AGB stars, PIEs can be triggered if a sufficiently high overshoot is considered. These events lead to low $^{12}$C/$^{13}$C ratios, high Li abundances, and no enrichment in s-elements. 
We find that the $2-3$\Msun\ AGB models experiencing a PIE can account for most of the observational features of J-type stars and AB2 grains. The remaining tensions between models and observations are (1) the low $^{14}$N/$^{15}$N ratio of some AB2 grains and of 2 out of 13 J-type stars, (2) the high $^{26}$Al/$^{27}$Al of some AB2 grains, and (3) the J-type stars with A(Li)~$<2$. Extra mixing mechanisms can alleviate some of these
tensions, such as thermohaline or rotation. 
}
{This work  highlights a possible match between AGB stellar models that undergo a PIE and J-type stars and AB2 grains.
To account for other types of carbon stars, such as N-type stars, PIEs should only develop in a fraction of solar-metallicity AGB stars. 
Additional work is needed to assess how the occurrence of PIEs depends on mixing parameters and initial conditions, and therefore to further confirm or exclude the proposed scenario.
}


\titlerunning{}

\authorrunning{A. Choplin et al. }

\maketitle

\begin{table*}[h!]
\scriptsize{
\caption{ Properties of J-type stars (first part of the table) and surface properties of our AGB models just after the PIE.   
\label{table:jstars}
}
\begin{center}
\resizebox{18.5cm}{!} {
\begin{tabular}{l|ccccccccccc} 
\hline
\hline
 Star  & $T_{\rm eff}$ & [M/H]& C/O & A(Li) & $^{12}$C/$^{13}$C & $^{14}$N/$^{15}$N & $^{16}$O/$^{17}$O & $^{16}$O/$^{18}$O   & [F/Fe] &  [Ba/Fe]& Ref. \\ 
\hline
WX Cyg   & 2940 & 0.3     & 1.02 & 4.4 &4.5& $>6$             &      $-$                     &      $-$               &$-$ &  0.0    & $1,2, 3$ \\ 
WZ Cas   & 3140 & 0.0     &1.03  & 4.8 & 4  & $-$                &$504 \pm 130$          &$1250 \pm 350$ & 0.95  & 0.4     &   $1,3,4,5$ \\ 
VX And    & 2890 & 0.05   & 1.78 & 2.6 & 6  & $900$            &770                            &  $650\pm 170$ &  0.04 & 0.25    &   $1,2,3,4,5$ \\ 
UV Cam   & 3350 & 0.2     & 1.07 & 3.0 & 4 & $>700$          &      $-$                      &    $-$                 &  $-$  &$-0.07$&  $1,2,3$ \\ 
Y CVn      & 2860 & 0.0      & 1.10 & 0.7 & 3  & $3200$          &$270 \pm 55$           & $>300$             &$<-0.86$& 0.3       &    $1,2,3,4,5$ \\ 
FO Ser     & 2600 & 0.1     & 1.10 & 1.2 &13 & $-$                 &         $-$                   &    $-$                 &  $-$&0.1           &  $1,3$ \\ 
BM Gem  & 3000 & 0.2     & 1.05  & 1.5& 9  &$1330\pm800$&         $-$                   &   $-$                   &$-$&$-0.2$      &  $1, 2,3$ \\ 
RY Dra    & 3010 & $-0.05$& 1.17 & 1.3& 2  & $-$                   &$385\pm250$          & $>200$             &$-0.51$&  0.1           & $1,3,4,5$ \\ 
RX Peg    & 2890 &0.4     & 1.10 &1.5 & 8  &$1800\pm1100$&     $-$                      &  $-$                   & $-$&  $-0.4$  & $1,2$ \\ 
V614 Mon& 2850 & $-0.1$& 1.02 & 1.3& 8  & $-$                   &       $-$                   &  $-$                  & $-$& 0.27      &  $1,3$ \\ 
V353 Cas & 2810 & 0.3    &1.10 & 2.7 & 7  & $2400$            &        $-$                  &    $-$                 & $-$&  $-$       &  $1,2,3$ \\ 
R Scl        & $-$ & $-$   &$-$   &  $-$ & 19&  $-$                  &    300                      &   500                 & 0.01 & $-$       &  $4,5$\\
T Lyr        & $-$ & $-$   &$-$   &  $-$ & 3.2&  $-$                  &    $450\pm 60$       &   $-$                 & $-0.04$& $-$       &  $4,5$\\
\hline
\hline
Model       &  &    &    &  & &                   &           &                & &             & \\
\hline
M1.0\_Zlow & 2489 &$-0.5$& 11  & 4.8 & 5.7   &9805&643&973& 0.30 & 0.13 &  $-$ \\ 
M2.0\_Zlow & 3160 &$-0.5$& 2.7 & 3.9 &  7.0  &3936&131&871& $-0.02$ & 0.02&  $-$\\ 
M3.0\_Zlow & 3649 &$-0.5$& 0.8 & 3.5 &  6.2  &4021&180&814& $-0.05$ & 0.01&  $-$\\ 
M1.0\_Zsun & 2290 & 0.0     & 7.2 & 5.4&  6.7  &4630&1456&709& 0.76 &  0.03&  $-$\\ 
M2.0\_Zsun & 2925 & 0.0    & 1.4  & 4.0 &  8.7 &2636&247&753& 0.05 & 0.02&  $-$\\ 
M2.0\_ZsunB$^{\star}$ & 3190 & $0.0$ & 1.0  & 2.5 & 11  & 2896 &292 & 919 & 0.18 & 0.00 &  $-$\\ 
M3.0\_Zsun & 3389 & 0.0    & 1.1  & 3.8 &  9.5 &3310&305&750& 0.00 & 0.01&  $-$\\ 
\hline
\end{tabular}
}
\end{center}
}
\normalsize{
{References}. 1 - \cite{abia00}; 2 - \cite{hedrosa13}; 3 - \cite{abia97}; 4 - \cite{abia17}; 5 - \cite{abia15}

$^{\star}$ this model was computed with thermohaline mixing (cf. Sect.~\ref{sect:li} for details). 

}
\end{table*}


\section{Introduction}
\label{sect:intro}

J-type stars \citep[named by][]{bouigue54} are a subclass of carbon stars (C/O~$>1$) that represent about 15\% of the carbon stars  of the Galaxy and Large Magellenic Cloud (LMC) \citep{abia00, morgan03}, and their origin is still largely unknown.  
Their metallicity is approximately solar and they are believed to have initial masses $M_{\rm ini} \lesssim 2-3$\Msun . They are likely on the (early) asymptotic giant branch (AGB) phase. Also, J-type stars have strong $^{13}$C isotopic features leading to $^{12}$C/$^{13}$C~$<20$ \citep[e.g.][]{lambert86, harris87, abia97, abia00, ohnaka99, abia02, abia17} and do not show overabundances of s-elements \citep[e.g.][]{abia00, hatzidimitriou03}. A significant fraction \citep[80~\% according to][]{abia20} were reported to be Li-rich. \cite{abia96} argued that Li and $^{13}$C may be produced by a similar mechanism.
To our knowledge, no radial-velocity variations attesting to the presence of a binary companion have been detected in J-type stars. 
Standard AGB models face difficulties in reproducing these stars, and particularly their low $^{12}$C/$^{13}$C ratio. 
\cite{sengupta13} proposed a binary-star formation channel involving re-accretion of nova ejecta onto a main sequence companion. Another possibility involves the merger of a helium white dwarf with a red giant \citep{zhang13}. 

Pre-solar silicon carbide (SiC) grains are microscopic solids that are believed to have been formed before the formation of the Solar System. These grains condensed around stars during their advanced evolutionary stages.
The AB grains form a subclass of SiC grains characterised by low $^{12}$C/$^{13}$C ratios ($\lesssim 10$) and a wide range of $^{14}$N/$^{15}$N ratios (between $\sim 30$ and $\sim 10000$). AB1 grains have subsolar $^{14}$N/$^{15}$N ratios ($<440$), while AB2 grains have supersolar $^{14}$N/$^{15}$N ratios \citep{liu17a}. 
AB2 grains show similar features to J-type stars, supporting the idea that they may have been formed by them \citep{liu17b}. 
Standard AGB models of different initial masses and metallicities are thought to be at the origin of several types of SiC grains \citep[mainstream, Y and Z; e.g.][]{zinner06}, but cannot account for the chemical features of AB grains.
Astrophysical sites that may manufacture such grains include core-collapse supernovae following proton ingestion in the He/C layers of the progenitors  \citep{pignatari15, liu17a}, and low-mass CO novae \citep{haenecour16}. 

In this Letter, we investigate the possibility that J-type stars and AB2 grains correspond to AGB stars that experienced proton ingestion events (PIEs) in the convection-driven thermal pulses (TP). 
Section~\ref{sect:models} presents our AGB models, while Sect.~\ref{sect:comp} is dedicated to their comparison with the various observational properties of J-type stars and AB2 grains. Conclusions are given in Sect.~\ref{sect:concl}.


\section{AGB models experiencing proton ingestion}
\label{sect:models}

\subsection{Physical inputs}

We consider initial masses at the zero-age main sequence (ZAMS) of $M_\mathrm{ini} =  1$, 2, and 3\Msun{} and metallicities of [Fe/H]~$= -0.5$ and $0.0$ (Table~\ref{table:jstars}), with the solar mixture of \cite{asplund09}.
The models are computed with the stellar evolution code STAREVOL \citep[][and references therein]{siess00, siess06, goriely18c}. 
We used the same input physics as detailed in previous papers \citep{choplin21, goriely21, choplin22b, choplin22a, martinet24, choplin24}.  
The mass-loss prescription of \cite{schroder07} is used until the end of core-helium burning, and we switch to  the \cite{vassiliadis93} formulation  during the AGB phase. 
During the PIE, a nuclear network of 1160 nuclei is used to follow the nucleosynthesis and the nuclear burning is coupled to the  transport equations. 
Overshoot mixing is included according to the prescription of \cite{goriely18c}. The overshoot diffusion coefficient, $D_{\rm over}$ is defined as
\begin{equation}
D_{\rm over} (z) = D_{\rm min} \, \times \, \left( \frac{D_{\rm cb}}{D_{\rm min}} \right)^{(1-z/z^{*})^{p}}
\label{eq:os18}
,\end{equation}
where $z^{*} = f_{\rm over} \, H_p \, \ln(D_{\rm cb}) / 2$ is the distance over which mixing occurs, $D_{\rm min}$ is the value of the diffusion coefficient at the boundary, $z=z^{*}$, and $p$ is a free parameter controlling the slope of the exponential decrease in $D_{\rm over}$ with $z$. 
We adopted the default values of $D_{\rm min}=1$~cm$^2$\,s$^{-1}$ and $p=1$. 
We considered overshooting only above the TP with $f_{\rm over} = f_{\rm top} = 0.10$ (for the M1.0\_Zlow and M2.0\_Zlow models) or $0.12$ (for the other models) to trigger a PIE, as previously studied in \citet[][see Sect.~\ref{sect:struc} for more details]{choplin24}.

\begin{figure}[t]
\centering
\includegraphics[width=1\columnwidth]{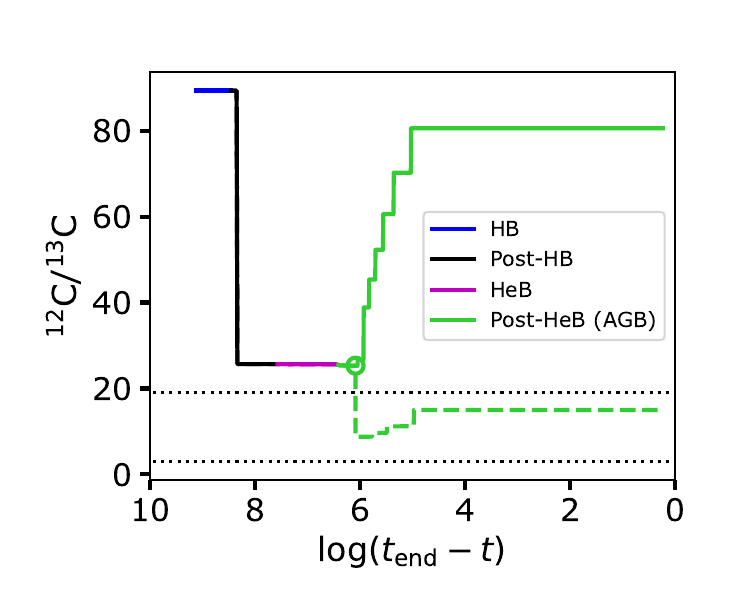}
\caption{Surface evolution of the $^{12}$C/$^{13}$C ratio in a 2\Msun\ model of solar metallicity. The different colours represent the different evolutionary stages (HB for core-hydrogen burning, HeB for core-helium burning). 
The solid line corresponds to the standard model computed without overshoot ($f_{\rm top} = 0$) and thus without PIE. 
The dashed line shows the same model but computed with $f_{\rm top} = 0.12$ and hence experiencing a PIE during the early AGB phase (denoted by the green circle).
The two dotted horizontal lines delineate the range of $^{12}$C/$^{13}$C ratios observed in J-type stars (Table~\ref{table:jstars}).}
\label{fig:ccsurf}
\end{figure}

\begin{figure*}[t]
\centering
\includegraphics[width=1.5\columnwidth]{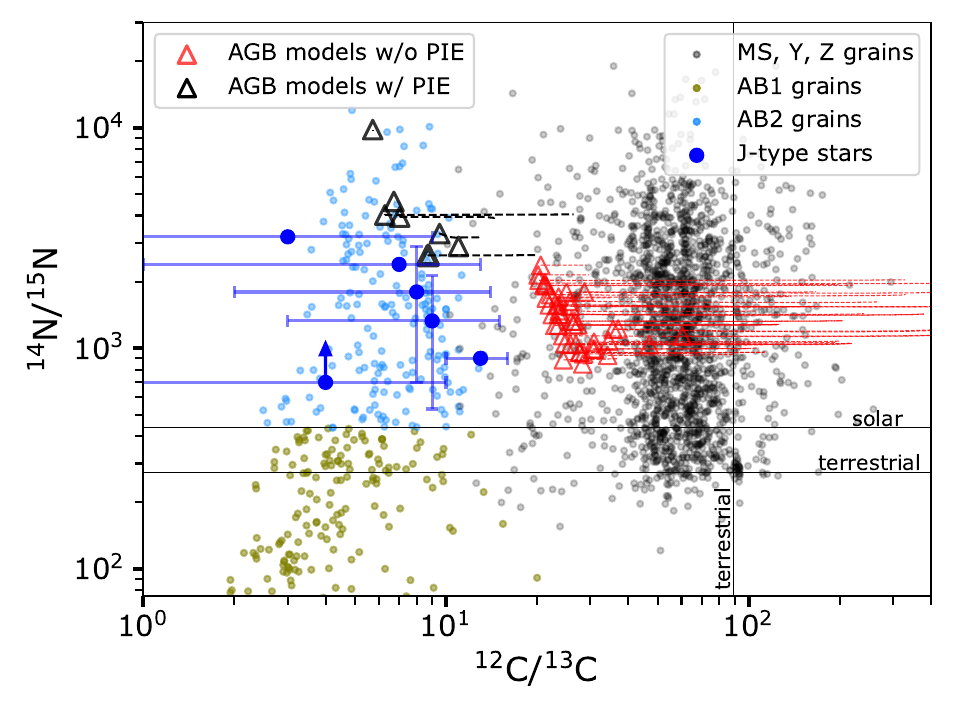}
\caption{$^{14}$N/$^{15}$N ratio as a function of $^{12}$C/$^{13}$C ratio in J-type stars (large blue points; also see Table~\ref{table:jstars}) and grains of various types (smal circles). Uncertainties on the $^{12}$C/$^{13}$C ratio in J-type stars amount to $\pm 6$, as recommended in \cite{abia00}. 
In red are shown the surface abundances of AGB models that do not experience PIEs, with initial masses and metallicities in the range of $1.3<M_\mathrm{ini}/M_\odot<6$ and $Z_{\odot}<Z<0.003$, corresponding to $-0.7 <$~[Fe/H]~$<0.0$ \citep[from the FRUITY database,][]{cristallo11, cristallo15}. The red triangles give the surface abundances after the first 3DUP and the red dashed lines correspond to the evolution along the AGB. 
Black triangles show the surface abundances of the AGB models computed in this work, just after the PIE. The thick black triangle corresponds to the solar-metallicity 2\Msun\ model (M2.0\_Zsun). Dashed black lines show the evolution of the surface abundances after the PIE.
Horizontal and vertical solid lines show solar and terrestrial $^{12}$C/$^{13}$C and $^{14}$N/$^{15}$N values for comparison.
}
\label{fig:ccnn}
\end{figure*}

\subsection{Structure and evolution}
\label{sect:struc}

The development of PIEs in AGB models has been extensively discussed in previous works \citep[e.g.][]{iwamoto04, cristallo09a, choplin21}. We reiterate just the main features of our models here, which are summarised in Table~\ref{table:jstars}. 
During a PIE, protons are engulfed by the growing convective TP, activating the reaction chain $^{12}$C($p,\gamma$)$^{13}$N($\beta^+$)$^{13}$C($\alpha,n$)$^{16}$O. Because of the local production of energy in the pulse by $^{12}$C($p,\gamma$)$^{13}$N, the convective pulse splits. A few years later, the upper part of the pulse eventually merges with the convective envelope, enriching the AGB surface with the products of the PIE nucleosynthesis.
Without extra mixing, PIEs are found to develop in AGB with $M_{\rm ini} < 2.5$\Msun\ and [Fe/H]~$<-2$ \citep{choplin24}. 
Including overshooting at the outer boundary of the TP facilitates PIEs. 
More specifically, the 1 and 2\Msun\ models with [Fe/H]~$=-0.5$ considered here experience PIEs when using $f_{\rm top} = 0.10$. 
During a PIE, the neutrons produced by $^{13}$C($\alpha,n$)$^{16}$O  reach  densities of $N_n \sim 10^{13-15}$~cm$^{-3}$ at the bottom of the TP. In our models, we have $3.1 \times 10^{12} < N_n < 8.2 \times 10^{13}$~cm$^{-3}$. 
At low-metallicity, the PIE leads to the copious production of trans-iron elements \citep[e.g.][]{cristallo09a, choplin21} but at solar metallicity, the synthesis of heavy elements is very modest and is concentrated around the iron-peak elements (Sect.~\ref{sect:heavy}).

After the PIE, the envelope is strongly enriched in metals, increasing its opacity and therefore the mass loss. The 1\Msun\ stellar models quickly lose their envelope, aborting the AGB evolution and preventing the development of further TPs. The model star of 2\Msun\ at  [Fe/H]~$=-0.5$ (0.0)  experiences 4 (2) additional TPs after the PIE, followed by third dredge-up (3DUP). 
For the 3\Msun\ with [Fe/H]~$=-0.5$ (0.0), 13 (16) pulses develop after the PIE, also followed by 3DUPs.


\section{Comparison to J-type stars and AB2 grains}
\label{sect:comp}

We consider the sample of 11 J-type stars of \citet{abia00} plus 2 additional J-type stars (R Scl and T Lyr) analysed in \cite{abia15, abia17}. Table~\ref{table:jstars} summarises the characteristics of these stars. 
For grains, we use the Presolar Grain Database\footnote{\url{https://presolar.physics.wustl.edu/presolar-grain-database/}} \citep[PGD,][]{hynes09, stephan24}. We select all AB2 grains with $^{14}$N/$^{15}$N~$>440$. 

\subsection{The C/O ratio}
\label{sect:co}

J-type stars have $1.02 <$~C/O~$< 1.78$ (Table~\ref{table:jstars}). This is compatible with the 2 and 3\Msun\ solar metallicity AGB models that have a surface C/O ratio within the range of $1 < $~C/O~$< 1.4$ just after the PIE (Table~\ref{table:jstars}). The 3\Msun\ at [Fe/H]~$=-0.5$ is also somewhat compatible with C/O~$=0.8$. Other models have too high C/O ratios.
After the PIE, further pulses and 3DUPs develop in the 2 and 3\Msun\ models (Sect.~\ref{sect:struc}). 
This increases the C/O ratios from 1.4 to 2.7 (1.1 to 1.5) in the M2.0\_Zsun (M3.0\_Zsun) model.
As a test, we recomputed the M2.0\_Zsun with overshoot above and below the pulse as well as below the envelope ($f_{\rm over} = 0.12$). In this case, two additional TPs develop after the PIE (as in the standard case) and the C/O ratio increases from 1.4 to 2.1. The lower final C/O is due to the overshoot mixing  below the pulse, which makes it hotter and leads to a higher oxygen intershell abundance \citep{herwig00}. 

\subsection{The $^{12}$C/$^{13}$C ratio}
\label{sect:cc}

Our models start with $^{12}$C/$^{13}$C~$=89$ (solar value). After the first dredge up, the surface $^{12}$C/$^{13}$C ratio drops to $20 - 30$ (Fig.~\ref{fig:ccsurf}), before rising again during the AGB phase (without PIE): the 3DUP events following the TPs bring $^{12}$C in the envelope and the $^{12}$C/$^{13}$C ratio steadily increases as the star climbs the AGB (solid green line in Fig.~\ref{fig:ccsurf}). 
Standard AGB models cannot account for $^{12}$C/$^{13}$C~$\lesssim 20$ (red lines in Fig.~\ref{fig:ccnn}) and consequently were proposed to be at the origin of MS, Y, and Z grains (Fig.~\ref{fig:ccnn}) that have larger $^{12}$C/$^{13}$C ratios. 
PIEs produce $^{13}$C mostly thanks to $^{12}$C($p,\gamma$)$^{13}$N($\beta^+$)$^{13}$C, which is later dredged up. The surface $^{12}$C/$^{13}$C ratio after the PIE in our AGB models is always between 5 and 10 (black triangles in Fig.~\ref{fig:ccnn}), which is compatible with the values determined in J-type stars and in AB2 grains. 
Further TPs (and 3DUPs) develop in our 2 and 3\Msun\ models and increase the $^{12}$C/$^{13}$C ratio up to about 20 at maximum (dashed black lines in Fig.~\ref{fig:ccnn}).

\subsection{The $^{14}$N/$^{15}$N ratio}
\label{sect:nn}

Immediately after the PIE, our models predict  $2636<^{14}$N/$^{15}$N~$<9805$ at the surface, which corresponds to the upper range of observed values in AB2 grains (Fig.~\ref{fig:ccnn}). 
The subsequent AGB evolution does not significantly alter this ratio. Out of the seven J-type stars with a $^{14}$N/$^{15}$N ratio (Table~\ref{table:jstars}), five are compatible with our models (within uncertainties) and two are below our model predictions, with ratios of 900 (VX And) and $1330 \pm 800$ (BM Gem). Slightly different initial masses and metallicity models may lead to lower $^{14}$N/$^{15}$N ratios. Also, additional mixing processes operating during the AGB phase and not included here may transport protons below the convective envelope and produce additional $^{15}$N during the interpulse via $^{14}$N($n,p$)$^{14}$C($\alpha,\gamma$)$^{18}$O($p,\alpha$)$^{15}$N, with the neutrons coming from $^{13}$C($\alpha,n$) \citep[see e.g.][]{goriely00, lugaro04, cristallo14, vescovi21}.

\subsection{Oxygen isotopic ratios}
\label{sect:oo}

J-type stars have $270<$~$^{16}$O/$^{17}$O~$<770$ (Table~\ref{table:jstars}). Within uncertainties, all stars (but VX And) are compatible with $^{16}$O/$^{17}$O~$= 300-400$. This is in agreement with our solar metallicity 2 and 3 \Msun\ models, which predict $^{16}$O/$^{17}$O~$\sim 300$. The $^{16}$O/$^{17}$O ratio tends to increase with decreasing mass (Table~\ref{table:jstars}), meaning that higher ratios (matching that of VX And) may be obtained in AGB models slightly less massive than 2\Msun.  
The predicted $^{16}$O/$^{18}$O ratios of $750 - 900$ in the solar metallicity 2 and 3 \Msun\ models are also compatible with those of J-type stars (within uncertainties). No oxygen isotopic ratios are available for AB2 grains.

\subsection{Fluorine}
\label{sect:fluorine}

Fluorine was determined in six J-type stars and is not enhanced except in WZ Cas, which has [F/Fe]~$=0.95$ (Table~\ref{table:jstars}).  We note that this star was classified as J-type by \cite{abia00} but as SC-type by, for example, \cite{abia15}. 
In our models, after the PIE, the surface [F/Fe] is close to zero except in M1.0\_Zsun, where [F/Fe]~$=0.76$ (Table~\ref{table:jstars}). This is compatible with the overall low values found in J-type stars.

\subsection{Lithium}
\label{sect:li}

\begin{figure}[t]
\centering
\includegraphics[width=0.9\columnwidth]{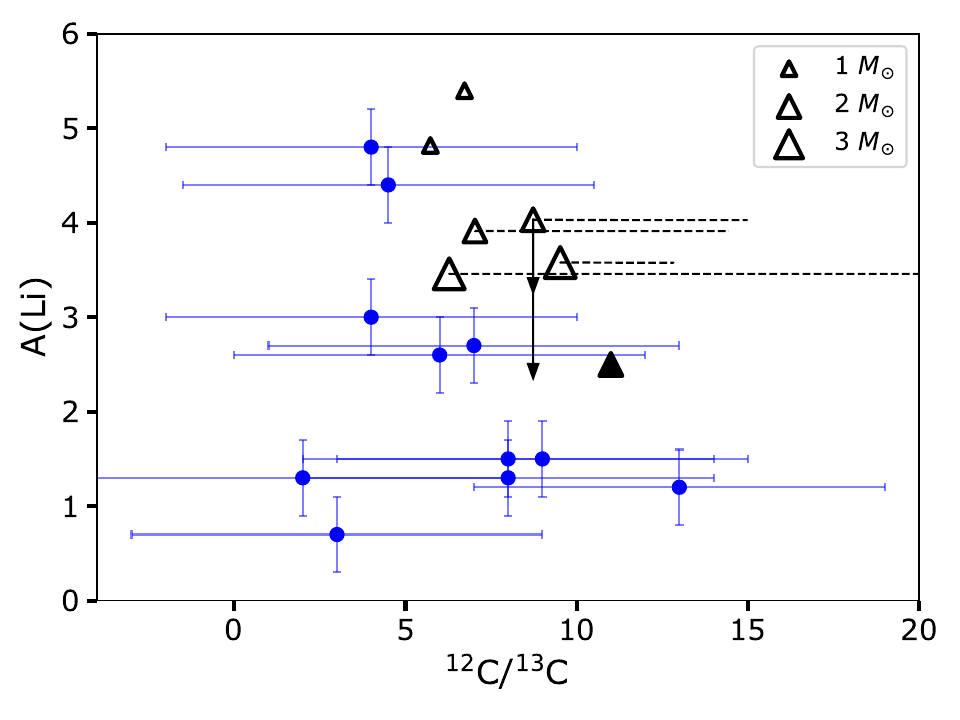}
\caption{
A(Li) as a function of the $^{12}$C/$^{13}$C ratio in J-type stars (blue dots, Table~\ref{table:jstars}) and at the surface of our AGB models immediately after the PIE (black triangles). Dashed black lines show the changes in abundances during the subsequent AGB evolution.
The short (long) vertical arrow indicates the lithium abundance if the $^{3}$He mass fraction in the H-rich envelope of the M2.0\_Zsun model is divided by 10 (100) before the PIE. 
The filled triangle corresponds to a 2\Msun{} AGB at solar metallicity computed with thermohaline mixing (cf. Sect.~\ref{sect:li} for details).  
}
\label{fig:ccli}
\end{figure}

Lithium is destroyed during the first dredge up and in our models at the start of the PIE, and the surface A(Li)~$\approx 0$.
During the PIE, $^{3}$He is engulfed in the pulse and burns via $^{3}$He($\alpha,\gamma$)$^{7}$Be (which is active at $T\gtrsim 40$~MK), later producing $^{7}$Li through the electron capture $^{7}$Be($e^-,\nu_e$)$^{7}$Li in the so-called \cite{cameron71} mechanism. The resulting surface A(Li) was shown by \citet{iwamoto04} to reach values of $3-5$ for $1-3$\Msun\ models with [Fe/H]~$=-2.7$.
In our models, we find $3.5<$~A(Li)~$<5.4$ after the PIE (Table~\ref{table:jstars}). This is compatible with about half of J-type stars (Fig.~\ref{fig:ccli}). The other half have A(Li)~$<2$.
The lower A(Li) in J-type stars may be the result of extra mixing processes below the convective envelope during the  thermally pulsating AGB phase (e.g. overshooting, thermohaline, or rotation) that allow the transport of the freshly synthesised Li to deeper and hotter regions where it is destroyed. 
An early depletion of $^{3}$He in the stellar envelope before the PIE would also hamper the Li production. Thermohaline and rotational mixing have been shown to decrease the envelope $^{3}$He before the AGB phase \citep[e.g.][]{charbonnel95, charbonnel07, eggleton08, lagarde11, lagarde12}.
We tested this scenario by artificially decreasing the $^{3}$He mass fraction by a factor of 10 and 100 in the envelope of our M2.0\_Zsun model just before the PIE. The resulting surface lithium enrichment (Fig.~\ref{fig:ccli}, black arrows) is reduced by $\approx 0.8$ and 1.8~dex, respectively. 
We also recomputed the M2.0\_Zsun model (named M2.0\_ZsunB in Table~\ref{table:jstars}) including thermohaline mixing following the scheme of \cite{kippenhahn80} and adopting a value of 1000 for the thermohaline coefficient, as in \cite{charbonnel07}. The resulting lithium abundance after the PIE is A(Li)=2.5 (filled triangle in Fig.~\ref{fig:ccli}), which is about 1.5~dex lower than the standard model.
Extra mixing processes can therefore account for some spread in the observed Li abundances.
We emphasise that our AGB models can reach the maximum A(Li) values obtained in J-type stars, which is a strong constraint.

\subsection{Aluminium}
\label{sect:al2627}

In the PSG, the $^{26}$Al/$^{27}$Al ratio was determined in about 50 AB2 grains and varies between  $4 \times 10^{-5}$ and $1.3 \times 10^{-2}$ with a mean value of $3 \times 10^{-3}$.
After the PIE, our AGB models predict $6.2 \times 10^{-5} <$~$^{26}$Al/$^{27}$Al~$<2 \times 10^{-3}$ at their surface, with a mean of $8.5 \times 10^{-4}$. 
The M2.0\_Zsun model has $^{26}$Al/$^{27}$Al~$=2 \times 10^{-3}$, which is compatible with the average $^{26}$Al/$^{27}$Al ratio in AB2 grains but cannot account for the highest values.
However, nuclear uncertainties associated with $^{26}$Al($p,\gamma$)$^{27}$Si, $^{25}$Mg($p,\gamma$)$^{26}$Al, $^{26}$Al($n,p$)$^{26}$Mg, and $^{26}$Al($n,\gamma$)$^{23}$Na can impact the production and destruction of $^{26}$Al  \citep[see e.g. discussion in][]{liu21}.

\subsection{Heavy elements}
\label{sect:heavy}

\begin{figure}[t]
\centering
\includegraphics[width=0.9\columnwidth]{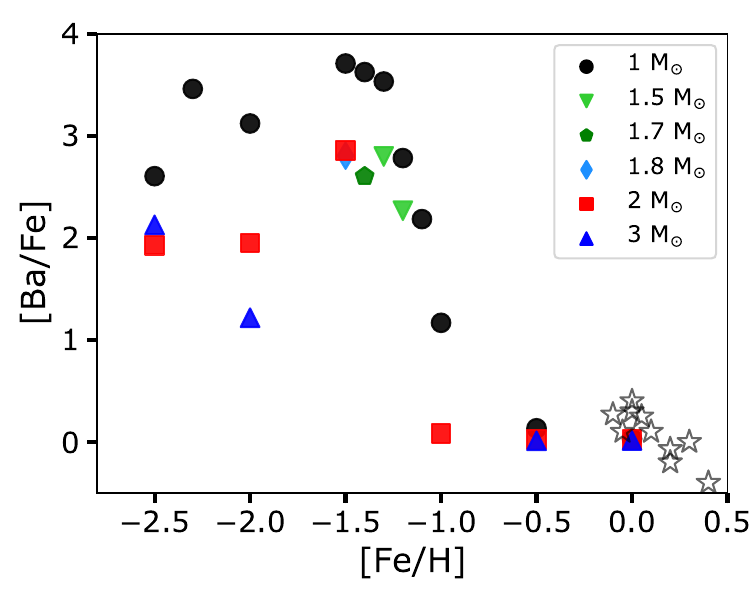}
\caption{Surface [Ba/Fe] ratio immediately after the PIE in AGB models with different masses and metallicities \citep[models from][plus those computed in the present work]{choplin24}. The grey star symbols represent J-type stars. 
}
\label{fig:bafe}
\end{figure}

The mean heavy-element enhancement in J-type stars is $0.13 \pm 0.12$, which is compatible with no enrichment \citep{abia00}.
Only low-metallicity AGB models with [Fe/H]~$\la -1$ experiencing a PIE during the early AGB phase are found to be strongly enriched in heavy elements by the i-process nucleosynthesis (Fig.~\ref{fig:bafe}).
For larger metallicities, the neutron-to-seed ratio during proton ingestion is too low to synthesise elements far beyond Fe (e.g. Barium).  
In particular, our solar-metallicity 2\Msun\, model does not show any overabundance of Ba, which is compatible with J-type stars (shown as black stars in Fig.~\ref{fig:bafe}). In this model, only elements with $26<Z<40$ are slightly enriched with respect to the Sun (0.3 dex at most). Including overshooting above and below the pulse and below the envelope in the M2.0\_Zsun model (cf. Sect.~\ref{sect:co}) leads to some s-processing after the PIE, with the surface [Ba/Fe] ratio rising from 0.02 after the PIE to 0.39 at the end of AGB (0.08 in the standard model).

\begin{figure*}[t]
\centering
\includegraphics[width=1.8\columnwidth]{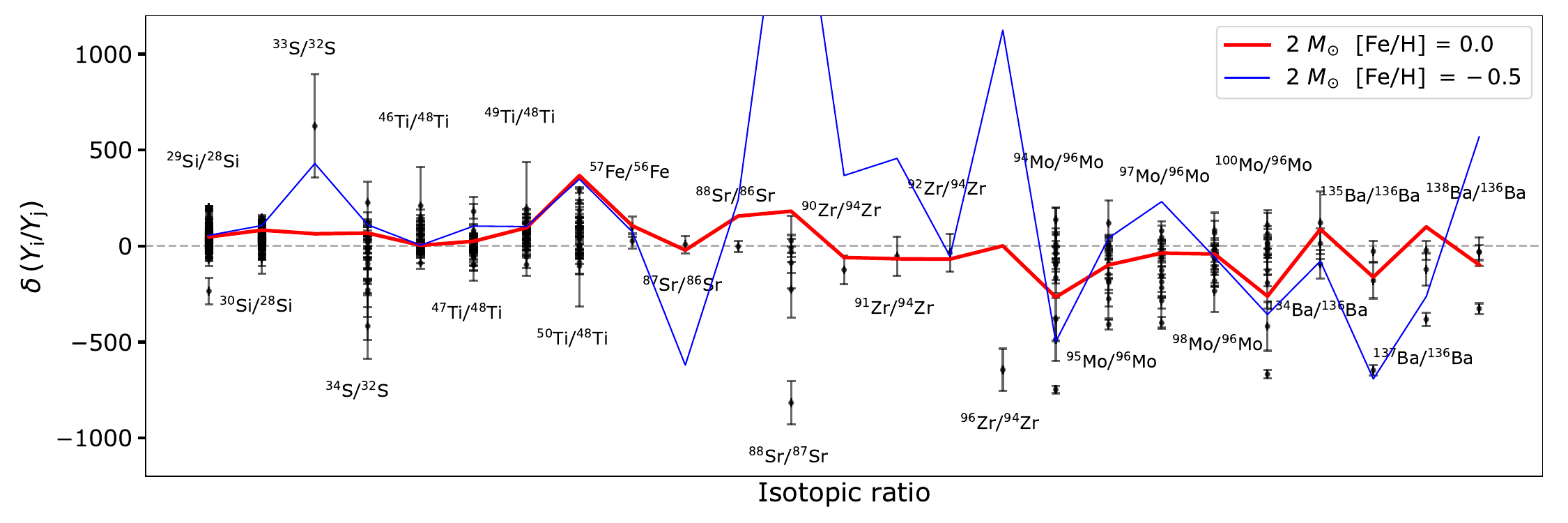}
\caption{$\delta (Y_{\rm i} / Y_{\rm j})$ ratios (in \textperthousand) defined by Eq.~\ref{eq:delta} for AB2 grains (i.e. AB grains with $^{14}$N/$^{15}$N~$>440$, black error bars) and two of our AGB models (Table~\ref{table:jstars}) that experienced a PIE (solid lines). 
}
\label{fig:isoAB2}
\end{figure*}

We compared the yields of two of our AGB models (M2.0\_Zsun and M2.0\_Zlow) to the available isotopic ratios extracted in AB2 grains. 
All the data available (using the PGD) for AB2 grains are shown in Fig.~\ref{fig:isoAB2} except $^{12}$C/$^{13}$C (cf. Sect.~\ref{sect:cc}), $^{14}$N/$^{15}$N (cf. Sect.~\ref{sect:nn}), and $^{26}$Al/$^{27}$Al (cf. Sect.~\ref{sect:al2627}), and a few ratios including p-isotopes not followed in our AGB models. 
If $Y_i, Y_j$ are the number abundances of isotopes $i$ and $j$, one can define the associated $\delta$-value, giving the permil deviations of $Y_i/Y_j$ with respect to standard ratios (typically representing solar values); it can be written as
\begin{equation}
\label{eq:delta}
\delta(Y_i/Y_j) = \left[ \frac{(Y_i/Y_j)_{\rm grain}}{(Y_i/Y_j)_{\rm standard}} - 1 \right] \times 1000~.
\end{equation}
As seen in Fig.~\ref{fig:isoAB2}, most ratios are well reproduced by the M2.0\_Zsun model but the comparison is not as good for the M2.0\_Zlow model. In particular, the large $^{88}$Sr/$^{87}$Sr ratio in this lower-metallicity model is due to the higher neutron-to-seed ratio during the PIE, leading to the synthesis of some neutron-richer isotopes like $^{88}$Sr, to the detriment of, for example, $^{87}$Sr.

\section{Summary, discussion, and conclusions}
\label{sect:concl}

In this Letter, we present an investigation of the possibility that J-type carbon stars are solar metallicity AGB stars ([Fe/H]~$ \approx 0$) that experienced a PIE and are the manufacturers of AB2 grains. 
Clearly, the most important features of J-type stars are their enrichment in $^{7}$Li and $^{13}$C and an absence of heavy elements. 
At solar metallicity, proton ingestion is a mechanism that co-produces $^{7}$Li and $^{13}$C, but no heavy elements. 
Overall, we find that most observational constraints of J-type stars and AB2 grains (metallicity, C/O ratio, lithium, fluorine, heavy elements, and isotopic ratios) can be accounted for by solar-metallicity $2-3$\Msun\ AGB models that experienced a PIE. 
A few tensions between models and observation nevertheless remain: (1) the $^{14}$N/$^{15}$N ratios below about 1500 in two J-type stars and in some AB2 grains cannot be accounted for by our models, (2) the high $^{26}$Al/$^{27}$Al in some AB2 grains cannot be reproduced, and (3) J-type stars with A(Li)~$<2$ cannot be explained. 
Tensions (1) and (2) could be alleviated by including extra mixing mechanisms, such as rotation or thermohaline mixing, acting before the PIE and/or during the subsequent AGB phase. 
We should stress that our surface-abundance predictions are subject to various uncertainties associated in particular with the mass-loss rate and the efficiency of the 3DUPs that will impact the number of TPs and therefore the envelope composition. With additional 3DUP episodes, the C/O ratio and s-process enrichment will increase, which is in contradiction with J-type stars properties. It might be that  J-type stars are a transient stellar population but further investigations are needed to clarify this point. 

If the PIE scenario is valid, it is unclear as to why J-type stars of only $\sim 2-3$\Msun\ are detected, and not $\sim 1$\Msun\ for instance. A simple reason might be that it takes too long for these metal-rich low-mass stars to reach the AGB, and provided that they can (if their mass is slightly higher), after the PIE these models quickly lose their envelope, reducing the duration of the AGB phase to about 0.03~Myr. By contrast, this phase lasts about 1.2 and 3.0~Myr in our 2 and 3~\Msun\ models, respectively, because of their more massive envelopes. In this scenario, the probability of observing J-type stars of $2-3$\Msun\ is consequently much higher. Finally, we note that PIEs are less likely to develop with increasing initial mass \citep[e.g.][]{choplin24}. This may explain the absence of J-type stars above $\sim 3$\Msun.

Importantly, for
N-type stars\footnote{N-type stars are another class of carbon stars showing
some s-process enrichment, higher $^{12}$C/$^{13}$C ratios, and no lithium
in general.} and MS grains to also be accounted for, which are expected to
originate from $\sim 2-3.5~M_{\odot}$ AGB stars following a standard evolution
\citep[i.e. no PIE; e.g.][]{abia20, abia22,straniero23}, only a fraction of solar-metallicity stars with initial masses of $\sim 2-3$\Msun\ can experience a PIE (and become J-type stars). 
As PIEs are more easily triggered in low-mass AGB stars, J-type stars should be less massive than N-type stars 
on average. This is somewhat consistent with the finding that J-type stars are dimmer (higher $M_{\rm bol}$) than N-type stars  on average \citep[Fig.~7 in][]{abia22}, although the overlap in $M_{\rm bol}$ is rather large.
We presently lack understanding as to why specific AGB stars experience a PIE while others do not. 
Our studies have shown that whether or not a PIE is triggered very much depends on the strength and location of the overshooting \citep{choplin24}. The occurrence of PIEs may also be impacted by other physical processes linked, for example, to rotational mixing \citep[e.g.][]{langer99, siess04, piersanti13, denhartogh19}, magnetic instabilities  \citep[e.g.][]{nucci14,denissenkov09,vescovi20}, or internal gravity waves \citep{Deni2003}; these waves were shown to impact AGB evolution but their effect on the development of PIEs has not yet been studied. 
Most of the chemical properties of J-type stars can be explained by 2-3\Msun\ stars experiencing a PIE but the reason why only a fraction of these stars experience this unusual mixing process remains unclear and further investigation is needed.

 \begin{acknowledgements}
A.C. is post-doctorate F.R.S-FNRS fellow.
L.S. and S.G. are senior F.R.S-FNRS research associates. 
 \end{acknowledgements}


\bibliographystyle{aa}
\bibliography{astro.bib}


\end{document}